\documentclass{article}
\usepackage{ijcai25}

\usepackage{times}
\usepackage{soul}
\usepackage{url}
\usepackage[hidelinks]{hyperref}
\usepackage[utf8]{inputenc}
\usepackage[small]{caption}
\usepackage{graphicx}
\usepackage{amsmath}
\usepackage{amsthm}
\usepackage{booktabs}
\usepackage{algorithm}
\usepackage{algorithmic}
\usepackage{amssymb}
\usepackage{mathtools}
\usepackage{threeparttable}
\usepackage[switch]{lineno}
\DeclareMathOperator{\sgn}{sgn}

\usepackage{color}

\newtheorem{definition}{Definition}

\newcommand{\h}{\mathbf{h}}
\newcommand{\p}{\mathbf{p}}
\newcommand{\q}{\mathbf{q}}

\title{Hierarchical Locality Sensitive Hashing for Structured Data: A Survey}

\author{
Wei Wu$^{1}$\and
Bin Li$^2$
\affiliations
$^1$School of Computer Science and Engineering, Central South University, Changsha 410083, China\\
$^2$School of Computer Science, Fudan University, Shanghai 200433, China\\
\emails
william.third.wu@gmail.com,
libin@fudan.edu.cn
}

\begin{document}

\maketitle

\begin{abstract}
Data similarity (or distance) computation is a fundamental research topic which fosters a variety of similarity-based machine learning and data mining applications. In big data analytics, it is impractical to compute the exact similarity of data instances due to high computational cost. To this end, the Locality Sensitive Hashing (LSH) technique has been proposed to provide accurate estimators for various similarity measures between sets or vectors in an efficient manner without the learning process. Structured data (e.g., sequences, trees and graphs), which are composed of elements and relations between the elements, are commonly seen in the real world, but the traditional LSH algorithms cannot preserve the structure information represented as relations between elements. In order to conquer the drawback, researchers have been devoted to the family of the hierarchical LSH algorithms. In this paper, we explore the present progress of research into hierarchical LSH from the following perspectives: 1) Data structures, where we review various hierarchical LSH algorithms for three typical data structures and uncover their inherent connections; 2) Applications, where we review the hierarchical LSH algorithms in multiple application scenarios; 3) Challenges, where we discuss some potential challenges as future directions.
\end{abstract}

\section{Introduction}


Data are growing prodigiously. For example, in 2024, approximately 2.2 trillion search queries are processed by all search engines each year\footnote{\url{https://www.businessdasher.com/search-engine-statistics}}; 
Big data have been driving machine learning and data mining research in both academia and industry. Data similarity (or distance) computation is a fundamental research topic which benefits numerous similarity-based machine learning and data mining applications, e.g., classification, clustering, retrieval, etc. However, it has been daunting for big data analytics to exactly calculate similarity due to the ``3V'' characteristics (i.e., volume, velocity and variety). A typical example is that it is intractable to conduct document analysis on the classic data mining book, Mining of Massive Datasets~\cite{rajaraman2012mining}, which contains over $10^8$ features in the case of 5-shingling, without data preprocessing, e.g., dimension reduction or feature selection. 
Many efficient algorithms working fluently in the low-dimensional spaces may run unacceptably slowly in the high-dimensional spaces 
due to high computational cost. Therefore, it is necessary to develop efficient yet accurate similarity estimation algorithms in the era of big data.

One powerful solution to addressing the above-mentioned problem is Locality Sensitive Hashing (LSH)~\cite{indyk1998approximate,broder1998min,gionis1999similarity,charikar2002similarity,datar2004locality,andoni2006near,weinberger2009feature}, which provides the unbiased estimators for some classical similarity (or distance) measures, e.g., MinHash for the Jaccard similarity~\cite{broder1998min}, Weighted MinHash for the generalized Jaccard similarity \cite{haveliwala2000scalable}, SimHash for the Cosine similarity~\cite{charikar2002similarity}, $p$-stable LSH for the $l_p$ distance~\cite{datar2004locality} 
and Feature Hashing for the inner product \cite{weinberger2009feature}. Intuitively speaking, an LSH algorithm maps similar data instances to the same hashcode with a higher probability than dissimilar ones by employing a set of randomized hash functions. No need of learning process is also a major reason that the LSH technique is highly efficient. Thus far researchers have widely applied the LSH technique to various real-world applications, e.g., social network analysis \cite{wu2023mpsketch}, bioinformatics \cite{tan2023schash}, high-energy physics application \cite{miao2024localitysensitive}, anomaly detection \cite{zeng2023double}, integrated circuit design \cite{chu2024attentionrc}, crash report detection \cite{remil2024deeplsh}, etc.

The above traditional LSH algorithms are designed for sets or vectors. However, in the real-world scenarios, there exist a variety of structured data, which are composed of elements and relations between the elements, e.g., genes, trajectories, parse trees, XML documents, molecules, social networks, protein-protein interaction networks, etc. From the perspective of data structures, they can be represented as sequences, trees or graphs. Furthermore, the structured data have widely solved many important machine learning problems such as classification, regression, clustering, trajectory mining, network analysis, etc. 
Obviously, these structured data are totally different from sets or vectors, and thus cannot be simply represented as sets or vectors and then fed into an traditional LSH algorithm because such a direct application may lose valuable relation information, which leads to severe performance loss. To capture structure information derived from relations between elements, a family of hierarchical LSH algorithms have been proposed to \emph{recursively} or \emph{iteratively} sketch the elements along the relation links.


A recent survey \cite{jafari2021survey} reviews the distributed LSH frameworks and the applications of LSH in diverse domains. By contrast, this survey emphasizes the hierarchical LSH algorithms for the structured data and its applications modeled as the structured data. We explore the hierarchical LSH algorithms from the following perspectives:
\begin{enumerate}
    \item \emph{Data Structures}: We introduce multiple hierarchical LSH algorithms\footnote{We have implemented some hierarchical LSH algorithms introduced in this survey at \url{https://github.com/AIandBD}.} for trees, sequences and graphs, and uncover their intrinsic connections -- The tree-structure naturally represents the hierarchical structures, and thus we employ tree-structures to decompose sequences or graphs and extract their intrinsic structure information.
    \item \emph{Applications}: We demonstrate the applications of the hierarchical LSH algorithms in a number of real-world application scenarios. 
    \item \emph{Challenges}: We propose some potential challenges w.r.t. the hierarchical LSH algorithms including explainability, heterogeneous structured data and acceleration of deep neural networks.
\end{enumerate}


\section{Preliminaries of LSH}
\label{sec:overview}

The nearest neighbor search problem aims to find the instance in a set that is closest to a given instance, which relies on data similarity (or distance) computation. Unfortunately, in big data analysis, it is impossible to return the exact solution within a reasonable time due to high computational overhead. For this reason, researchers resort to approximate solutions.

\begin{definition}[Approximate Near Neighbor~\cite{indyk2004nearest}]
  Given a set $\mathcal{P}$ on $N$ points in a metric space $M=(\mathcal{X}, dist)$, design a data structure that supports the following operation: For any query $\q\in\mathcal{X}$, if there exists $\p\in\mathcal{P}$ such that $dist(\p,\q)\le R$, find a point $\p'\in\mathcal{P}$ such that $dist(\p',\q)\le cR$, where $c>1$.
\end{definition}

A powerful solution to similarity (or distance) estimation is Locality Sensitivity Hashing (LSH), which exploits a family of randomized hash functions to map the similar data instances to the same data points in the low-dimensional space with a higher probability than the dissimilar ones. Based on such concise representation without the learning process, we acquire data similarity in an effective and efficient way and further conduct the similarity-based tasks. Formally, an LSH algorithm is defined as follows:
\begin{definition}[Locality Sensitivity Hashing~\cite{indyk1998approximate}]
  Given a family of hash functions $\mathcal{H}$, the probability of two instances, $\q$ and $\p$, to have the same hashcode by $h\in\mathcal{H}$ is exactly equal to their similarity\footnote{There is a more common definition for LSH in the language of $(R, cR, p_1, p_2)$-sensitivity; but we adopt the alternative definition of LSH as an unbiased estimator here for easier understanding.}:
\begin{equation}
    \mathbb{E}_h[h(\q)=h(\p)]=sim(\q,\p)\nonumber
\end{equation}
\end{definition}

In the following we provide brief introduction to the representative LSH instances which are building blocks for the hierarchical LSH algorithms reviewed in this survey.

\begin{definition}[MinHash~\cite{broder1998min}]
  Given a universal set $\mathcal{U}=\{u_1,\cdots,u_N\}$, a subset $\mathcal{S}\subseteq \mathcal{U}$ and a set of $K$ random permutations from the uniform distribution, $\{\sigma_k|\sigma_k: \mathcal{U} \mapsto \mathcal{U}\}_{k=1}^{K}$, the elements in $\mathcal{S}$ which are placed in the first position of each permutation would be the $K$-dimensional MinHash-codes of $\mathcal{S}$, $\h=\{\arg\min(\sigma_k(\mathcal{S}))\}_{k=1}^{K}$.
\label{minwise hashing}
\end{definition}

MinHash gives an unbiased estimator for the Jaccard similarity of two sets $Jaccard(\mathcal{S},\mathcal{T})=\frac{|\mathcal{S} \cap \mathcal{T}|}{|\mathcal{S} \cup \mathcal{T}|}$, and thus we have 
\begin{equation}\label{eq:min}
\Pr[\arg\min(\sigma(\mathcal{S}))=\arg\min(\sigma(\mathcal{T}))] = Jaccard(\mathcal{S},\mathcal{T}).\nonumber
\end{equation}

MinHash places each element in the first position with equal probability, which leads to serious information loss in the case of weighted sets because the weights denoting the importance of each element are simply replaced with 1 or 0. To this end, Weighted MinHash~\cite{wu2018review} was proposed to estimate the generalized Jaccard similarity of the weighted sets, $Jaccard_g(\mathcal{S},\mathcal{T})=\frac{\sum_{i}\min(s_i,t_i)}{\sum_{i}\max(s_i,t_i)}$, where $s_i$ and $t_i$ are the weights, by placing each element in the first position with the probability in proportion to its weight. 

\begin{figure*}[t]
\centering
\includegraphics[width=\linewidth]{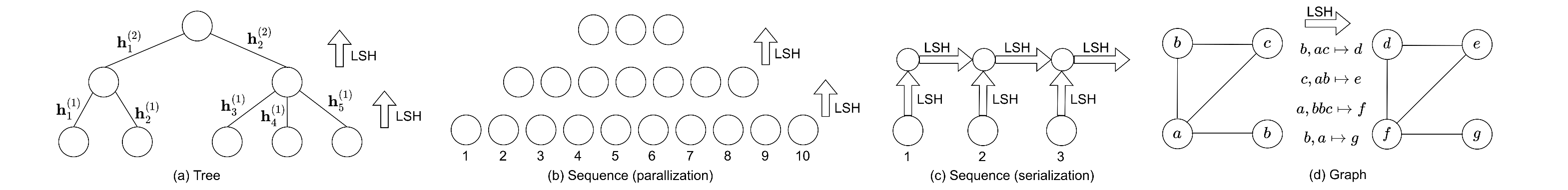}
\caption{An illustration of hierarchical LSH on structured data.}
\label{fig:structured_data}
\vspace{-10pt}
\end{figure*}

\begin{definition}[SimHash~\cite{charikar2002similarity}]
  Given a vector $\mathbf{u}$ and a set of $K$ random hyperplanes $\{\mathbf{r}_k\}_{k=1}^{K}$ drawn from the Gaussian distribution, $\{h_{\mathbf{r}_k}|h_{\mathbf{r}_k}(\mathbf{u})=\sgn(\mathbf{r}_k\cdot\mathbf{u})\}_{k=1}^{K}$, where $\sgn(\cdot)=1$ if $\mathbf{r}_k\cdot\mathbf{u}\ge 0$; $\sgn(\cdot)=0$ otherwise, would return the $K$-dimensional SimHash-codes of $\mathbf{u}$, $\h=\{h_{\mathbf{r}_k}(\mathbf{u})\}_{k=1}^{K}$.
\end{definition}
SimHash unbiasedly estimates the Cosine similarity of two vectors, $Cosine(\mathbf{u}, \mathbf{v}) = 1-\frac{\arccos\frac{\mathbf{u}\cdot\mathbf{v}}{\|\mathbf{u}\|_2\|\mathbf{v}\|_2}}{\pi}$. We have
\begin{equation}\label{eq:sim}
\Pr[h_{\mathbf{r}}(\mathbf{u})=h_{\mathbf{r}}(\mathbf{v})] = Cosine(\mathbf{u}, \mathbf{v}).\nonumber
\end{equation}

\begin{definition}[Feature Hashing~\cite{weinberger2009feature}]
  Given a vector $\mathbf{u}=[u_1,\cdots,u_N]$, two randomized hash functions $h:\mathbb{N}\mapsto\{1,2,\cdots,K\}$ and $\xi:\mathbb{N}\mapsto\{\pm1\}$, the $K$-dimensional hashcode of $\mathbf{u}$ is defined as  $\h=\varphi^{(h,\xi)}(\mathbf{u})$, where $h_k=\varphi^{(h,\xi)}_k(\mathbf{u})= \sum_{j:h(j)=k}\xi(j)u_j$.
\end{definition}
This algorithm gives an unbiased estimator for the inner product of two vectors $\mathbf{u}\cdot\mathbf{v}$, and thus we have
\begin{equation}\label{eq:inner}
    \mathbb{E}[\langle\varphi^{(h,\xi)}(\mathbf{u}),\varphi^{(h,\xi)}(\mathbf{v}) \rangle]=\mathbf{u}\cdot\mathbf{v}.\nonumber
\end{equation}

\begin{definition}[$p$-stable LSH~\cite{datar2004locality}]
  Given a vector $\mathbf{u}$ and a set of $K$ random vectors $\{\mathbf{a}_k\}_{k=1}^{K}$ sampled from $p$-stable distribution ($p\in(0,2]$), $\{h_{\mathbf{a}_k, b}|h_{\mathbf{a}_k, b}(\mathbf{u})=\lfloor\frac{\mathbf{a}_k \cdot \mathbf{u}+b}{w}\rfloor\}_{k=1}^{K}$, where $b\sim \text{Uniform}(0,w)$ and $w$ is the length of the interval when the real axis is segmented into the equal-width intervals, would return the $K$-dimensional $p$-stable hashcodes of $\mathbf{u}$, $\h=\{h_{\mathbf{a}_k, b}(\mathbf{u})\}_{k=1}^{K}$.
\end{definition}
This algorithm solves the problem under $l_p$ norm based on $p$-stable distributions, where $p\in(0,2]$. We have
\begin{equation}
    \Pr[h_{\mathbf{a}, b}(\mathbf{u})\!\!\!=\!\!\!h_{\mathbf{a}, b}(\mathbf{v})]=\int_0^w\dfrac{1}{\|\mathbf{u}-\mathbf{v}\|_p}f_{p}(\dfrac{t}{\|\mathbf{u}-\mathbf{v}\|_p})(1-\dfrac{t}{w})dt,\nonumber 
\end{equation}
where $f_p$ is the probability density function of the absolute value of the $p$-stable distribution. Particularly, it is the Cauchy distribution for $p=1$ and the Gaussian distribition for $p=2$.


Also, some classic algorithms, e.g., L1LSH \cite{gionis1999similarity} and L2LSH \cite{andoni2006near}, are designed for the $l_1$ and $l_2$ distances, respectively.

The above basic LSH algorithms are designed for sets or vectors, so they consider the elements independently and cannot extract the relations between the elements if they exist, but they have paved the way for the hierarchical LSH algorithms introduced in the following.

\section{Data Structures}
\label{sec:structured}

The structured data, e.g., texts, genes, trajectories, time series, molecules and social networks, are very common, and have been widely applied to document analysis, bioinformatics, network analysis, etc. They contain both information from elements and the structure information constituted by the relations between the elements, and thus cannot be simply represented as sets or vectors because the structure information has to be thrown away in this case. A great many hierarchical LSH approaches have been proposed to address the issue. 
In this section, we roughly categorize the structured data into \emph{trees}, \emph{sequences} and \emph{graphs}. Note that a sequence is a special case of a tree, and each node of a graph can be naturally represented as a rooted tree. Therefore, these three commonly seen types of structured data can be all hierarchically hashed based on the tree structure with the LSH technique. Table \ref{tab:hierarchical_lsh} summarizes the hierarchical LSH algorithms.


\begin{table*}[t]
\setlength{\abovecaptionskip}{5pt}%
\setlength{\belowcaptionskip}{0pt}%
  \centering
  \begin{threeparttable}
  \caption{An overview of hierarchical LSH algorithms} \label{tab:hierarchical_lsh}
  \begin{normalsize}
  \fontsize{7.5pt}{\baselineskip}\selectfont{\begin{tabular}{lrrrrr}
  \hline\noalign{\smallskip}
  Hierarchical LSH Algorithms & Data Structures	& Basic LSH Schemes	& Similarity/Distance & Time Complexity \tnote{1} & Applications \\
  \noalign{\smallskip}\hline\noalign{\smallskip}
  \cite{gollapudi2008power} & Tree	& MinHash	& Jaccard & $\mathcal{O}(\sum_{r=1}^{R}|\bigcup\h_i^{(r-1)}|^rK)$ & Document Analysis\\
  \cite{chi2014context} &Tree	&MinHash	&Jaccard & $\mathcal{O}(\sum_{r=1}^{R}|\h_i^{(r-1)}|^rK^{R-r+1})$ & Document Analysis \\
\noalign{\smallskip}\hline\noalign{\smallskip}
 Gallager LSH \cite{luo2016low} & Sequence	& L2LSH	&$l_2$ distance & $\mathcal{O}(MK4^{q'})$ & Bioinformatics \\
 SLSH \cite{kim2016stratified} &Sequence	&SimHash, L1LSH	& Cosine,~$l_1$ distance & $\mathcal{O}(MK^2)$ & Healthcare \\
 POISketch\cite{yangpoisketch} &Sequence	& Weighted MinHash	& Generalized Jaccard	& $\mathcal{O}(MK+d)$ \tnote{2} & Urban Computing \\
 HistoSketch \cite{yang2017histosketch} &Sequence	& Weighted MinHash	&Generalized Jaccard	& $\mathcal{O}(MK+d)$ \tnote{2} & Urban Computing \\
 MRTS \cite{astefanoaei2018multi} &Sequence	& - & Fr\'echet distance & $\mathcal{O}(M\log KR+n)$ \tnote{3} & Urban Computing \\
 D$^2$HistoSketch \cite{yang2018d} &Sequence	& Weighted MinHash	& Generalized Jaccard	& $\mathcal{O}(MK+d)$ \tnote{2}  & Urban Computing \\
 \noalign{\smallskip}\hline\noalign{\smallskip}
 NSH \cite{li2012nest} & Graph & Feature Hashing & Inner Product & $\mathcal{O}(|\mathcal{E}|R)$ & Bioassay Testing \\
 NetHash \cite{wu2018efficient} & Graph & MinHash & Jaccard  & $\mathcal{O}(|\mathcal{V}|K^2\mathrm{e}^{-(2R-1)S}\overline{\nu}^R)$ & Social Network Analysis \\
 KATH \cite{wu2017k} & Graph & MinHash & Jaccard  & $\mathcal{O}(|\mathcal{V}|R)$ & Bioassay Testing \\
 NodeSketch \cite{yang2019nodesketch} & Graph & Weighted MinHash & Generalized Jaccard  & $\mathcal{O}(|\mathcal{V}|K(R-2)\overline{\nu}^2)$ & Social Network Analysis  \\
 \#GNN \cite{wu2021hashing} & Graph & MinHash & Jaccard & $\mathcal{O}(|\mathcal{V}|KR(K+\overline{\nu}))$ & Social Network Analysis \\
 SGSketch \cite{yang2022streaming} & Graph  & Weighted MinHash & Generalized Jaccard  & $\mathcal{O}(|\mathcal{V}|K(R-2)\overline{\nu}^2)$ & Social Network Analysis  \\
 NodeSig \cite{celikkanat2022nodesig} & Graph  & 1-stable LSH & $\chi^2$ similarity & $\mathcal{O}(|\mathcal{E}|RK)$  & Social Network Analysis  \\
 ELPH \cite{chamberlain2023graph} & Graph & MinHash & Jaccard & $\mathcal{O}(|\mathcal{E}|R|\mathcal{A}|+KR^2+R|\mathcal{A}|^2)$ & Social Network Analysis \\
 SCHash \cite{tan2023schash} & Graph & SimHash & Cosine & $\mathcal{O}(R\sum_{c=0}^C|\mathcal{R}_c|^3K^3)$ \tnote{4} & Bioassay Testing \\
 LoNeSampler \cite{kutzkov2023lone} & Graph & MinHash & Jaccard & $\mathcal{O}(|\mathcal{V}||\mathcal{A}|+|\mathcal{E}|R)$ & Social Network Analysis \\
 MPSketch \cite{wu2023mpsketch} & Graph & MinHash & Jaccard & $\mathcal{O}(|\mathcal{V}|K^2R\overline{\nu})$ & Social Network Analysis \\
 \noalign{\smallskip}\hline\noalign{\smallskip}
  \end{tabular}
  }
  \tiny{
  \begin{tablenotes}
    \item[1] We simplify time complexity due to space limit.
    \item[2] As shown in the original papers, $d$ is the number of rows in count-min sketch.
    \item[3] As shown in the original papers, $n$ is the total number of disks intersecting $\mathcal{T}$.
    \item[4] As shown in the original papers, $C$ is the dimensionality of simplicial complex and $|\mathcal{R}_c|$ is the number of $c$-dimensional simplexes.
  \end{tablenotes}}
  \end{normalsize}
  \end{threeparttable}
  \vspace{-10pt}
\end{table*}

\subsection{Trees}
\label{subsec:trees}

A tree composed of nodes and directed edges describes the hierarchical structure by the directed edges from the parent node to the child nodes. Furthermore, if the nodes are assigned with content information, the tree will have stronger semantic capability. The tree structure is commonly used to represent the hierarchical structure of a text.

As shown in Figure \ref{fig:structured_data}(a), LSH can recursively sketch a tree representing a top-down hierarchy of chapters, paragraphs and sentences from the leaf nodes to the root node, which generates the $K$-dimensional hashcodes for each node at each level. Obviously, such a representation in Eq.~(\ref{eq:text}) is richer than the ordinary bag-of-words model, 
\begin{eqnarray}
  \mathcal{S}^{(r)}&=&\{\h_i^{(r-1)}|i=1,2,\cdots,|\mathcal{S}^{(r)}|\},\label{eq:text}
\end{eqnarray}
where $\h$ is the representation of each node in the tree and $r\in\{1,2,\cdots,R\}$ denotes the $r$th iteration. The text information in the leaf nodes are propagated along the edges from bottom to top. The hashing process can be formally expressed as 

\begin{equation}
    \h^{(r)} = \operatorname{LSH}^{(r)}(\h_1^{(r-1)},\h_2^{(r-1)},\cdots,\h_{|\mathcal{S}^{(r)}|}^{(r-1)}).
    \label{eq:tree}
\end{equation}

\cite{gollapudi2008power} directly concatenates the hashcodes in Eq.~(\ref{eq:text}) as the flat set, while \cite{chi2014context} reorganizes them into the nested set with $K$ set-elements, each of which represents the features extracted from all the nodes at this level. The nested set saves more context information, because the nested set with $K$ set-elements results from $K$ sampling operations on all the nodes at the lower level and then they will be independently processed. Consequently, \cite{chi2014context} can compare the internal nodes in probability and propagate lower-level similarities to higher levels for the sake of better performance.

Tree-based hierarchical LSH algorithms effectively and efficiently capture the relations between elements, which further lays the foundation of sequence and graph data hashing.


\subsection{Sequences}

A sequence is a set of elements that appear in a specific order $\mathcal{T}=(T_1,T_2,\cdots, T_M)$, e.g., genes, trajectories and time series. The elements in a sequence are hierarchically hashed to preserve the order information in a parallel or serial manner. 

From the perspective of parallelization, the data at the higher level, which are treated as the set data, are encoded as the small-sized data at the lower level, as shown in Figure \ref{fig:structured_data}(b). Formally, they can be described as
\begin{equation}
    \h^{(r)} = \operatorname{LSH}^{(r-1)}(\h^{(r-1)}),
    \label{eq:trajectory}
\end{equation}
where $\h$ is the representation of the sequence, and $r$ is the $r$th iteration.

The common sequence data, e.g., texts and genes, generally adopt the technique of $q$-shingling in NLP and $q$-mers in bioinformatics to extract features. Longer $q$-mers mean to preserve more dependency within large contexts and lead to higher accuracy. However, it substantially increases time and space cost because huge training sets are required. Gallager LSH \cite{luo2016low} is constructed hierarchically by employing a small number of L2LSH functions in Eq.~(\ref{eq:trajectory}) to capture the structure information and represent long $q$-mers as shorter $q'$-mers level by level. 


Time-series data and trajectories have multi-facet information, e.g., frequency, amplitude, trend and coordinates. 
SLSH \cite{kim2016stratified} explores time-series data with more diverse and refined perspectives by building a two-level LSH framework, where multiple distance metrics are supported at different levels. Specifically, in Eq.~(\ref{eq:trajectory}), the outer-level L1LSH coarsely stratifies the data via amplitude under the $l_1$ distance; then the inner-level SimHash finely sketches each stratified data via angle and shape under the Cosine distance. Consequently, SLSH preserves much more information through different similarity measures in a hierarchical way than the traditional LSH algorithms, since the latter provides only one perspective on the data with its associated similarity measure. 


Fr\'{e}chet distance is popular in trajectory comparison since it incorporates the inherent sequential characteristics. The dynamic programming algorithms for Fr\'{e}chet distance have $O(MN)$ time complexity, where $M$ and $N$ are the lengths of two compared trajectories, respectively. To lower the time complexity and make it feasible for comparing large-scale trajectories, MRTS \cite{astefanoaei2018multi} develops a disk-based LSH algorithm whose main insight is that when disks are deployed randomly, similar trajectories will be likely to intersect a similar set of disks. Based on that, MRTS hierarchically sketches trajectories and directly estimates their distance. Specifically, it adopts the random disks with different sizes at different levels in Eq.~(\ref{eq:trajectory}). 

From the viewpoint of serialization, the data at the present moment are incrementally encoded based on the previous result, as shown in Figure \ref{fig:structured_data}(c). The mode is formally as follows
\begin{equation}
    \h^{(r)} = \operatorname{LSH}^{(r-1)}(\h^{(r-1)}, T_r),
    \label{eq:serial}
\end{equation}
where $\h$ is the representation of the sequence and $T_r$ denotes the emerging data at the $r$th moment.

Streaming histogram data $\mathbf{T}=[T_1,T_2,\cdots]$ is also a type of trajectory data, where $T_r$ denotes the cumulative count of the
$r$-th histogram element. A certain element arrives and then the corresponding count is updated at each time moment. POISketch \cite{yangpoisketch} instantiates LSH via Weighted MinHash in Eq.~(\ref{eq:serial}) and incrementally updates the hash value of the changed element. Note that the concept drift issue in the streaming data seriously impairs the accuracy of similarity computation. To address this issue, based on POISketch, HistoSketch \cite{yang2017histosketch} and D$^2$HistoSketch \cite{yang2018d} introduce the forgetting mechanism \cite{gama2014survey} -- the weights denoted by the counts of the elements exponentially decay as time goes by, while the hash values of the unchanged elements can be scaled proportionally according to the mathematical property of Weighted MinHash. 




\subsection{Graphs}
\label{subsec:graphs}


A graph $G=(\mathcal{V},\mathcal{E}, f_v, f_e)$ is composed of the node set $\mathcal{V}$, the edge set $\mathcal{E}$ and two mappings $f_v: \mathcal{V}\mapsto \mathcal{A}_v$ and $f_e: \mathcal{E}\mapsto \mathcal{A}_e$, which assign attributes from the attribute sets $\mathcal{A}_v$ and $\mathcal{A}_e$ to each node and edge, respectively. Also, we let $\overline{\nu}$ and $S$ be the average degree and the entropy of the graph, respectively. Particularly, $\mathcal{A}_v/\mathcal{A}_e=\emptyset$ indicates no attributes on the nodes/edges. Two nodes $v,u\in\mathcal{V}$ are neighbors if they are connected by an edge $e_{vu}\in\mathcal{E}$. 
Graph data are very ubiquitous in our daily life, e.g., molecules and social networks.

The Weisfeiler-Lehman (WL) graph kernel \cite{shervashidze2011weisfeiler} is a classic graph comparison algorithm based on the WL isomorphism testing \cite{weisfeiler1968reduction}, which extracts subtrees as features from the graph. The core step is to iteratively update the node information by aggregating information from its own and all its neighbors. We find an interesting connection between the WL graph kernel and some early hierarchical LSH algorithms for graph data, and propose a general graph hashing framework shown in Figure \ref{fig:structured_data}(d), which also resembles the Message Passing Neural Networks (MPNN) \cite{gilmer2017neural} by replacing nonlinear activation functions in MPNN with LSH functions. Specifically, the LSH-based works generate the weight matrices in a way of randomness not learning. Formally, the framework is composed of aggregation operation $\operatorname{AGG}$, message function $\operatorname{MSG}$ and node update function $\operatorname{UPD}$, 
\begin{eqnarray}
      m_v^{(r)} &=& \operatorname{AGG}_{u\in \mathcal{N}(v)}(\operatorname{MSG}^{(r-1)}(\h_u^{(r-1)})), \label{eq:message}\\
      \h_v^{(r)} &=& \operatorname{UPD}^{(r-1)}(\h_v^{(r-1)}, m_v^{(r)}),\label{eq:update}
\end{eqnarray}
where $\h_v$ represents node $v$, $r$ is the $r$th iteration, $\operatorname{AGG}$ and $\operatorname{MSG}$ aggregates and transforms information from the neighbors, respectively; $\operatorname{UPD}$ returns the latest node embedding by updating information from the node and all its neighbors. 


To our best knowledge, NSH \cite{li2012nest} is the first hierarchical LSH algorithm for graph data. It adopts the identity function for $\operatorname{MSG}$, instantiates $\operatorname{UPD}$ via Feature Hashing, and adopts the concatenation operation on the node and all its neighbors as $\operatorname{AGG}$, which significantly improves the WL graph kernel in time and space. However, it suffers from clear performance loss. KATH \cite{wu2017k} was proposed to achieve the competitive performance as the WL graph kernel with dramatically reduced computation and storage. In KATH, $\operatorname{MSG}$ employs truncation/filling or MinHash to store a varying number of neighbors in fixed-sized vectors, $\operatorname{AGG}$ is implemented by the concatenation operation on the node and all its corresponding neighbors, and $\operatorname{UPD}$ is instantiated as MinHash. In order to improve efficiency, KATH fast obtains the neighbors of nodes by introducing the traversal table with the fixed number of columns. Furthermore, SCHash \cite{tan2023schash} explores the higher-order interactions between nodes in the graphs instead of node pairwise interaction by introducing a kind of higher-order abstraction -- simplicial complexes. Specifically, SCHash first transforms node-based graphs into simplicial complex-based graphs by extracting the simplicial complexes, and builds the Hodge Laplacian matrix based on the simplicial complexes as the adjacency matrix; subsequently, it adopts the identity function for $\operatorname{MSG}$, instantiates $\operatorname{UPD}$ via SimHash, and adopts the addition operation on the simplicial complex and all its neighbors as $\operatorname{AGG}$.


\begin{figure*}[t]
\setlength{\abovecaptionskip}{5pt}%
\setlength{\belowcaptionskip}{0pt}%
\centering
\includegraphics[width=\linewidth]{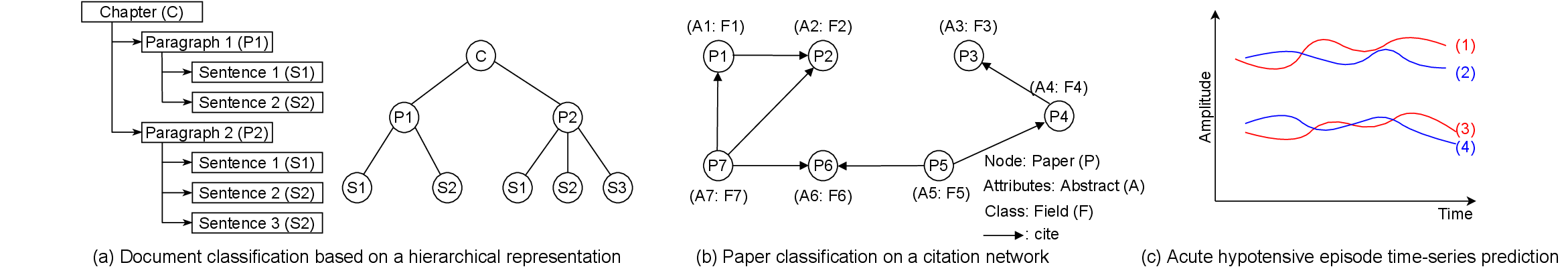}
\caption{Some examples of hierarchical LSH applications.}
\label{fig:applications}
\vspace{-10pt}
\end{figure*}

NetHash \cite{wu2018efficient} extends each node as a rooted tree along the edges in $\operatorname{AGG}$, and then recursively sketches the rooted tree from bottom to top. Within the framework, $\operatorname{MSG}$ is instantiated as MinHash on the neighbors, $\operatorname{AGG}$ aggregates the content information of the node and all its corresponding neighbors via the union operation, and finally $\operatorname{UPD}$ is implemented by MinHash on the aggregated results from the node and all its corresponding neighbors. Although NetHash significantly improves efficiency of exploring low-order neighbors at the expense of accuracy, its time complexity grows exponentially with the depth of the rooted tree because each rooted tree is independently sketched, which makes it hard to utilize high-order proximity. Similarly, LoNeSampler \cite{kutzkov2023lone}, \#GNN \cite{wu2021hashing} and MPSketch \cite{wu2023mpsketch} adopt the same operations as NetHash. LoNeSampler extends the direct neighborhood into multi-hop neighborhood and then just aggregates some important neighbors. \#GNN iteratively sketches each node and all its neighbors. In each iteration, all node information is shared in the whole graph for the next iteration, which makes the time complexity linear w.r.t. the number of iterations and in turn solves the problem of scalability. By contrast, MPSketch preserves more information than \#GNN by exploring a larger aggregated information pool. ELPH \cite{chamberlain2023graph} incorporates the node features with the structural features such as triangle counts by adopting MinHash and HyperLogLog \cite{flajolet2007hyperloglog}, which is the hashing technique and estimates the size of the union of the sets, in $\operatorname{MSG}$, the $\min$ and $\max$ operation in $\operatorname{AGG}$ and a learnable function in $\operatorname{UPD}$. Consequently, ELPH is more expressive than MPNN because it can distinguish the automorphic nodes which have the identical structural roles.

The above algorithms can capture content and structure information in the graph simultaneously. In contrast, NodeSketch \cite{yang2019nodesketch} preserves both the 1st and the 2nd order node structural proximities by adding the self-loop-augmented adjacency vectors of the graph, and recursively sketches each node and all its neighbors via the Weighted MinHash algorithm. Specifically, it implements $\operatorname{MSG}$ and $\operatorname{UPD}$ via Weighted MinHash, and conducts the union operation on the node and all its corresponding neighbors and the empirical distribution of information from the corresponding neighbors in $\operatorname{AGG}$. Inspired by HistoSketch, based on NodeSketch, SGSketch \cite{yang2022streaming} preserves the dynamic edge information in $M$ with the help of the forgetting mechanism \cite{gama2014survey} ---- the edge weights in the self-loop-augmented adjacency vectors exponentially decay as time goes by. Motivated by $p$-stable LSH, NodeSig \cite{celikkanat2022nodesig} implements $\operatorname{MSG}$ via the Cauchy random matrix, and then takes the sum of all the neighbors weighted by the transition probability that the node itself randomly walks towards the neighbors in $\operatorname{AGG}$ and the $\operatorname{sgn}$ operation in $\operatorname{UPD}$ on the aggregated result.

\subsection{Comparison with Representation Learning}

Representation learning aims to embed the complex data objects into a low-dimensional feature space, while accurately preserving similarity and laying the foundation for the downstream tasks such as classification. In recent years, Neural Networks (NN), which show the powerful representational ability in learning the hidden patterns, have been widely applied to representation learning in the above data structures. Despite the fact that the NN framework has achieved great success in various areas, it suffers from serious computational overhead and memory costs due to the expensive parameter learning involved, which hinders the application of these algorithms without the help of a powerful workhorse.

The LSH technique is a powerful tool to efficiently estimate the similarity of high-dimensional data objects by sketching the similar data objects as the same hashcode with a higher probability than the dissimilar ones in a non-learning manner. Therefore, it has been also used for representation learning, which strikes an excellent balance between accuracy and efficiency in the whole process. Consequently, The LSH technique seeks to remedy the gap in accuracy by virtue of a significant speedup, particularly for large-scale datasets.

\subsection{Theoretical Analysis}

Theoretically, the work \cite{morris2019weisfeiler} has proved that the WL isomorphism testing has the maximal representational power to distinguish the substructures in the graph because there is no mapping from different substructures to the same point in the low-dimensional feature vector. Actually, this theory can be extended to the structured data in this paper, and the representational power of the algorithm closely relies on the ability to discriminate the substructures in the structured data. As demonstrated in \cite{wu2021hashing}, the probability of the hierarchical LSH algorithms mapping two substructures into the same location equals their similarity. 

Based on one of the Weighted MinHash algorithms \cite{li20150}, POISketch, HistoSketch and D$^2$HistoSketch approximate the generalized Jaccard similarity with the same error bound as \cite{li20150}, but the further proof and its associated error bound remains a difficult problem. \cite{gollapudi2008power} and MRTS just provide the theoretical lower bound of the similarity and upper bound of the dissimilarity from the viewpoint of the definition. Considering the fact that we practically adopt multiple LSH functions to approximate the expected similarity, NSH, NetHash, LoNeSampler, \#GNN, MPSketch, NodeSketch, SGSketch, SCHash and \cite{chi2014context} further give the bounds of the deviation between the similarity estimator and the real similarity in terms of probability via the Hoeffding \cite{hoeffding1963probability}, Chernoff \cite{chernoff1952measure} or Bernstein \cite{bernstein1924modification} inequalities.

\section{Applications}
\label{sec:applications}

In this section, we introduce the applications of the hierarchical LSH algorithms in some real-world application scenarios. 

\subsection{Document Analysis}

LSH was initially applied to large-scale document analysis, i.e., web deduplication in the case of over 30,000,000 documents from the Aka Vista search engine \cite{broder1997resemblance}, where each document is represented as a set (bag-of-words representation). However, the hierarchical structure information, which comprises chapters, paragraphs and sentences, in the document cannot be captured, and thus it is hard to understand the context information. To this end, the document is organized into a tree structure to represent the above hierarchy \{chapters, \{paragraphs, \{sentences\}\}\}. As shown in Figure \ref{fig:applications}(a), the context information, for example, Paragraphs 1 and 2 contain two and three sentences, respectively, is embedded into the tree structure. \cite{gollapudi2008power,chi2014context} recursively sketch the tree-structured document as the low-dimensional hashcodes for classification. 

\subsection{Social Network Analysis}

Social networks are very common in our daily life. For example, in an academic community, a citation network shown in Figure \ref{fig:applications}(b) is composed of papers as nodes, where words in the abstract denote attributes of a node, and citations between papers as edges, and one can conduct academic paper classification via node classification; in World Wide Web, a social network consists of users with many profiles as nodes and friendships between users as edges, which can recommend friends to the users by link prediction. These applications can be implemented through network embedding, which represents each node in a network as a low-dimensional vector with the similarity between the nodes preserved. 
To this end, NetHash, NodeSketch, LoNe Sampler, \#GNN, MPSektch, SGSketch, NodeSig and ELPH recursively sketch each node and all its neighbors in a much more efficient manner than the GNN framework.

\subsection{Bioinformatics}
\label{subsec:bioinformatics}

Bioinformatics is an interdisciplinary research field aiming to understand biological data through mathematical, statistical and computer science tools. Gene analysis, most of which depends on similarity computation, is an important problem in bioinformatics. A gene is an enumerated collection of nucleotides (i.e., A, T, C and G) in which repetitions are allowed and order matters. In order to preserve the order information, the technique of $k$-mers is adopted, which are the substrings of length $k$ (similar to shingling in NLP). Consequently, the gene is transformed into a set of all its subsequences of length $k$, where each $k$-mer preserves the order information. Gallager LSH hierarchically sketches the set of $k$-mers as smaller-sized ones from bottom to top.

\subsection{Bioassay Testing}

A molecule is represented as a graph $G=(\mathcal{V},\mathcal{E}, \ell)$, where the atoms form the node set $\mathcal{V}$, the bonds between atoms form the edge set $\mathcal{E}$ and $\ell:\mathcal{V}\mapsto\mathcal{L}$ is a function that assigns a label (i.e., the name of the atom) from a label set $\mathcal{L}$ to each node. Also, a molecule is associated with a class label $y$, which denotes the result of the molecule in the bioassay test or clinical diagnosis, e.g., inflammasome and cancers. Molecular classification predicts the unseen molecules by learning a classifier from $\{G_i, y_i\}_{i=1}^{N}$. NSH, KATH and SCHash hierarchically sketch the subtrees composed of the nodes or the simplicial complexes and all their corresponding neighbors in each molecule, which is represented as the hashcode with similarity preserved as the kernel for graph classification.

\subsection{Urban Computing}

Urban computing solves problems in the urban development, e.g. traffic flows, points of interest and moving trajectories, by computer science, which helps to understand the essence of the urban phenomena and even predict the urban futures. One of the core research problems is to learn knowledge from spatio-temporal data, e.g. trajectories. Generally, a trajectory is represented as a sequence of ordered points $\mathcal{T}=(T_1, T_2, \cdots)$. MRTS estimates the distance between trajectories by hierarchically sketching the trajectories. 
POISketch, HistoSketch and D$^2$HistoSketch incrementally update the representation of the trajectories modelled as streaming histogram data in an iterative way.

\subsection{Healthcare}

Healthcare refers to maintenance or improvement of health with the help of sensors. Particularly, physiological time series observed in the monitors demonstrate the critical body status, particularly in Intensive Care Units (ICU). Physiological time series contain multiple human physiological indices from diverse perspectives, for example, acute hypotensive episode can be described as amplitude in mean blood pressure, and shape in trend and cycle frequency. Figure \ref{fig:applications}(c) shows that (1, 2) and (3, 4) are grouped as similar in the $l_1$ distance while (1, 3) and (2, 4) are grouped as similar in the Cosine distance under normalization. SLSH can preserve such different similarity measures by hierarchically sketching the data with a two-level LSH algorithm.

\section{Challenges}
\label{sec:challenges}

Although the hierarchical LSH family has fast developed with many practical implementations of algorithms, 
there are still a number of challenges for future exploration.

\subsection{Explainability}

Since the killing of a technician by an industrial robot in Germany, the EU has required that the algorithms present their principles of output \cite{bundy2019explainable}. Explainability in algorithms has drawn more and more attention because the explainable results help to grasp advantages and disadvantages, particularly in some high-risk scenarios, e.g., finance, healthcare, transport, etc. Unfortunately, the hash functions in LSH actually mix some features as a single one in the low-dimensional representation, which makes it hard to explain. 
Only few works \cite{wu2018efficient,wu2021hashing,wu2023mpsketch} hierarchically passes messages from each node to all its neighbors in the network from the viewpoint of sampling ---- the hash value is interpreted as a feature sampled from the node feature vectors, but they cannot still conduct qualitative analysis. It would be a great challenge to develop explainable, hierarchical LSH.

\subsection{Heterogeneous Structured Data}

Variety is one of the ``3V'' characteristics in big data, which means heterogeneity of data types. For example, a citation network covers heterogeneous information about authors, papers, conferences, terms, venues, etc. The heterogeneous features with distinct statistical properties are in totally different feature spaces, where inconsistency makes similarity between heterogeneous data hard to compute. Nevertheless, to the best of our knowledge, most of the existing LSH algorithms are designed for the homogeneous data, and the topic of the LSH technique for heterogeneous structured data is almost untouched ---- StreamHash \cite{manzoor2016fast} essentially put some heterogeneous features together as new, homogeneous features in the bag-of-words model. 

The key point is how to map the heterogeneous features into a common feature space. One promising approach is the Asymmetric Locality Sensitive Hashing (ALSH) \cite{shrivastava2014asymmetric}, which potentially represents the heterogeneous features in the common feature space by introducing different (i.e., asymmetric) transformation functions. Therefore, the design of asymmetric transformations is an interesting challenge and the key to combating with the problem.

\subsection{Acceleration of Neural Networks}

Neural Networks (NN) including Large Language Models (LLM) possess strong representational power to uncover intricate patterns within data. However, the current trend of having models with billions or more parameters can result in significant computational inefficiencies. The field of pure LSH-based representation learning for structured data is still in its infancy, and it deserves to further remedy the accuracy gap between the two classes of methods. \cite{dasgupta2017neural} discovers the connection between NN and LSH -- fruit fly olfactory circuit assigns similar neural activity patterns to similar odors, which suggests the potential of integrating LSH into the NN framework to leverage their respective advantages. To our best knowledge, some recent works \cite{meisburger2023bolt,daghaghi2021accelerating,liu2022halos} have adopted LSH to sample the major parameters, which constitute only a small proportion of the total parameters, and then maintain their updating frequency, while decreasing the frequency of minor parameter updates, because the number of parameters substantially impacting the NN methods is limited. Exploring the idea deeply could bring great insights to the NN and emerging LLM research.

\section{Conclusion and Discussion}
\label{sec:conclusions}

In this brief survey, we review the emerging hierarchical LSH algorithms for structured data including trees, sequences and graphs, and point out that these structured data can be decomposed into (sub)tree-structures for hierarchically hashing, which motivates the LSH community to design hierarchical LSH methods. Also, we introduce some typical applications from diverse fields and discuss the potential challenges.


Although the LSH-based representation learning for structured data have achieved success in a variety of application fields, their application is still limited compared to the NN-based methods. A possible reason is that the LSH-based algorithms largely rely on constructing the proper LSH functions with the favorable mathematical properties, in contrast to learning model parameters through optimizing the objective functions. This motivates us to write this survey, which introduces the LSH-based representation learning algorithms into the community and advocates for this alternative highly-efficient learning-free approach.

\bibliographystyle{named}
\bibliography{main}

\end{document}